\documentclass[letterpaper,12pt]{article}   
\usepackage{osajnl2} 
\usepackage[draft]{hyperref} 

\begin{document}

\title{Voltage-Controlled Negative Refractive Index in Vertically Coupled Quantum Dot Systems}
\author{Huan Wang and Ka-Di Zhu}
\address{Department of Physics, Shanghai Jiao Tong University,
Shanghai 200240, People's Republic of China}
\address{$^*$Corresponding author: wanghuan2626@sjtu.edu.cn}

\begin{abstract}
We demonstrate that voltage-controlled negative refractive index can
be obtained in self-organized InAs quantum dot systems. As the bias
voltage is changed, the refractive index can be adjusted and
controlled continuously from negative to positive and simultaneously
the loss of light in the system will be small. The single-negative
index materials and the double-negative index materials can be
achieved only by simply applying the external bias voltages.
\end{abstract}


\maketitle 
Far from a material with a negative index of refraction suggested by
Veselago in 1968\cite{Veselago}, the unique physical and good
optical properties such as the development of a "perfect
lens"\cite{Pendry} in which imaging resolution in not limited by
electromagnetic wavelength have attracted lots of interest. Early
experiments\cite{Shelby,Smith} which achieved the negative index are
some special structures with circuits and coils at microwave
frequencies. Recently, some nanoscale periodic structures including
photonic metamaterials demonstrate the negative index effect
\cite{Dolling}. Khoo\cite{Khoo} et al. have shown that aligned
nematic liquid crystal cells containing core-shell nanospheres are
possible to devise a new type of metamaterial with negative index.
Valentine et al. \cite{Valentine} experimentally realized the
three-dimensional optical metamaterial with a negative refractive
index. At the same time, negative index materials (NIM) are studied
theoretically by many researchers \cite{Oktel,Shen} in atomic
systems. Recently, properties of quantum dots (QD) in the optical
field are studied \cite{zhu,Jiang}. In this letter, we propose a
sheme consisting of a typical InAs self-assembled quantum dot system
and show that negative index of refraction becomes accessible only
simply by changing the bias voltage. The refractive index can be
adjusted continuously from negative to positive as the bias voltage
alters, simultaneously the loss of light in the system is very
small. In such a way the single-negative index materials (SNIM) and
the double-negative index materials (DNIM) can be easily achieved in
different bias voltages.

A vertically coupled InAs/InGaAs asymmetrical quantum dot system
consisting of two layers of dots (the upper layer and the lower
layer) with different band structures is shown in Fig.1(a). Samples
are arrays of InAs dots in a matrix of InGaAs which are vertically
stacked and electrically coupled in the growth direction. Dots in
two different layers show a strong tendency to align vertically. The
substrates and the top surface are composed with transparent and
conductive materials. $V$ is a bias voltage. The coupling of two
layers QD is mainly determined by the separation distance of two
layers. Fig.1(b) is a TEM picture of such self-assembled quantum
dots\cite{Qianghua}. In this quantum dot system, the lower QD is
slightly small, so its energy difference between ground state and
first excited state is larger than that of the upper one. From
Ref.~\cite{Krenner}, we know that for QD separation $d>9$ nm the
tunneling coupling between the two dots is weak and the QD system
can be discussed in terms of a simplified single-particle
picture\cite{zhu,Bester}. Applying an electromagnetical field we can
excite one electron from the valence to the conduction band in the
lower dot which can in turn tunnel to the upper dot.  Fig.1(c) shows
this transport. Using this configuration the Hamiltonian of the
system in the electric dipole approximation reads as follows
\cite{Villas}
\begin{equation}
H=\sum_{i=1}^{3} \hbar \omega_i |i><i| -\frac{\hbar}{2}(\Omega
e^{-i\nu t}|1><2| +\Omega^{*}e^{i\nu t}|2><1|)+\hbar
T_c(|1><3|+|3><1|),
\end{equation}
where $\hbar\omega_i$ is the energy of level $i$ of the quantum
dots, and $|i><j|$ $(i,j=1,2,3)$ are projection operators. $\nu$ is
the frequency of the laser.  Inter-dot tunneling between two layers
is described by a single, real parameter $T_c$ conventionally
\cite{Brandes}. The Rabi frequency associated with the optical
transition ($|2>\rightarrow|1>$) is
$\Omega=\overrightarrow{d_{12}}\cdot \overrightarrow{E}/\hbar$ where
$\overrightarrow{E}$ stands for the complex amplitude of the
positive frequency  component electric field of the laser and
$\overrightarrow{d_{12}}=e<1|\overrightarrow{r}|2>$ is the electric
dipole operator.

The density matrix elements of the system evolve according to the
Liouville equation \cite{Scully},
\begin{equation}
\frac{d\rho}{dt}=-\frac{i}{\hbar}[H, \rho]-\frac{1}{2}\{\Gamma,
\rho\}.
\end{equation}
In what follows we assume a diagonal relaxation matrix
$<i|\Gamma|j>=\gamma_i \delta_{ij}$, and choose the relaxation rates
for off-diagonal element of the density matrix as
$2\gamma_{ij}=\gamma_i+ \gamma_j$, and then
\begin{equation}
\dot{\rho_{12}}=-(i\omega_{12}+\gamma_{12})\rho_{12}+\frac{i}{2}\Omega
e^{-i\nu t}(\rho_{11}-\rho_{22})-iT_c\rho_{32},
\end{equation}
\begin{equation}
\dot{\rho_{32}}=-(i\omega_{32}+\gamma_{32})\rho_{32}+\frac{i}{2}\Omega
e^{-i\nu t}\rho_{31}-iT_c\rho_{12}.
\end{equation}
As the lower quantum dot is initially in the ground state $|2>$,
\begin{equation}
\rho_{22}^{0}=1,\qquad \rho_{11}^{0}= \rho_{33}^{0}=\rho_{13}^{0}=0.
\end{equation}
On substituting these initial conditions into (3) and (4), and
making the substitutions,
\begin{equation}
\rho_{12}=\tilde{\rho}_{12}e^{-i\nu t}, \qquad
\rho_{32}=\tilde{\rho}_{32}e^{-i\nu t}.
\end{equation}
After the substitution, this set of equation can be solved, for
example, by first writing in the matrix form \cite{Scully},
\begin{equation}
\dot{R}=-MR+A
\end{equation}
with
\begin{displaymath}
\mathbf{R} = \left( \begin{array}{ccc}
\tilde{\rho}_{12}  \\
\tilde{\rho}_{32}
\end{array} \right),
\end{displaymath}
\begin{displaymath}
\mathbf{M} = \left( \begin{array}{ccc}
\gamma_{12}+i\Delta & iT_c \\
iT_c & \gamma_{32}+i\delta \\
\end{array} \right),
\end{displaymath}
\begin{displaymath}
\mathbf{A} = \left( \begin{array}{ccc}
\frac{i}{2}\Omega \\
0
\end{array} \right),
\end{displaymath}
where $\Delta=\omega_{21}-\nu$ and
$\delta=\omega_{32}-\nu=\Delta-\omega_{31}$. Then integrating
\begin{equation}
R(t)=\int_{-\infty}^t e^{-M(t-t')}Adt'=M^{-1}A.
\end{equation}
We yield
\begin{equation}
\tilde{\rho}_{12}=\frac{\frac{i}{2}\Omega(\gamma_{32}
+i\delta)}{(\gamma_{12} +i\Delta)(\gamma_{32} +i\delta)+T_c^2},
\end{equation}
and
\begin{equation}
\tilde{\rho}_{32}=\frac{\frac{i}{2}\Omega(-iT_c)}{(\gamma_{12}
+i\Delta)(\gamma_{32} +i\delta)+T_c^2}.
\end{equation}
From Eq.(9) and Eq.(10), we have
\begin{equation}
\tilde{\rho}_{32}=-\frac{iT_c}{(\gamma_{32}+i\delta)}\tilde{\rho}_{12}.
\end{equation}
Therefore, the electric polarizability $\chi_{e}$ is given by
\begin{equation}
\chi_{e}(\omega)=\frac{2Nd_{12}\tilde{\rho}_{12}}{\varepsilon_{0}E}=\frac{i
N d_{12}^2}{\varepsilon_0
\hbar}\frac{\gamma_{32}+i\delta}{(\gamma_{12}+i\Delta)(\gamma_{32}+i\delta)+T_c^2},
\end{equation}
where $N$ is the density of the double quantum dots. The magnetic
susceptibility $\chi_m$ is expressed as
\begin{equation}
\chi_{m}(\omega) =\frac{2N m_{32}
\tilde{\rho}_{32}}{H_m}=-\frac{m_{32}}{d_{12}
c}\sqrt{\frac{\mu_r}{\varepsilon_r}}\frac{iT_c}{(\gamma_{32}+i\delta)}\chi_e(\omega),
\end{equation}
where $m_{32}$ is the magnetic dipole matrix element and $c$ is the
speed of light in vacuum. In Eq.(13), we have used the relation
$H_m=\sqrt{\frac{\varepsilon_r \varepsilon_0}{\mu_r \mu_0}}E$
between the envelopes of magnetic and electric fields, where
$\varepsilon_r =1+\chi_e$ and  $\mu_r =1+\chi_m$, and then we have
\begin{equation}
\epsilon_r =1+\chi_e=1+\frac{i Nd_{12}^2}{\varepsilon_0
\hbar}\frac{\gamma_{32}+i\delta}{(\gamma_{12}+i\Delta)(\gamma_{32}+i\delta)+
T_c^2},
\end{equation}
\begin{equation}
\mu_r=1+\frac{\eta^2\pm\sqrt{\eta^4+4\eta^2}}{2},
\end{equation}
where
\begin{equation}
\eta=-\frac{m_{32}}{d_{12}
c}\frac{iT_c}{(\gamma_{32}+i\delta)}\frac{\chi_e}{\sqrt{1+\chi_e}}.
\end{equation}
The results for the real and imaginary parts of the relative
electric permittivity $\epsilon_r$ and the relative magnetic
permeability $\mu_r$ as a function of $T_c$ with three different
detunings ($\Delta/\gamma_{12}=0.001, 0.002, 0.003$) are depicted in
Figure 2. In the figure, we consider $N=5\times10^{20}m^{-3}$,
$|m_{32}/(d_{12}c)|\sim10^{-2}$,  $\gamma_{12}\sim10^{9} s^{-1}$
\cite{Becher}, $\gamma_{32}\sim5\times10^{-4}\gamma_{12}$,
$d_{12}\sim50 Debye$ \cite{Kamada}, $\omega_{31}=0$ and
$\delta=\Delta$. From Fig.2 (a) and (b), we can see that the curves
of the real and imaginary part of $\epsilon_r$ are symmetrical when
$T_c$ varies from -0.5 to 0.5. Fig.2 (c) and (d) show that the real
and imaginary parts of $\mu_r$ are almost symmetrical. For
$\Delta=0.003\gamma_{12}$ and $T_c=\pm 0.2\gamma_{12}$, the real
parts of relative conductivity and permittivity are both negative.
This means a DNIM is achieved. In other cases,
$Re(\epsilon_r)<0,Re(\mu_r)>0$ or $Re(\epsilon_r)>0,Re(\mu_r)<0$,
one can achieve a SNIM. We also demonstrate the influence of
detuning on the relative conductivity and permittivity. As the
detuning $\Delta$ increases from $0.001\gamma_{12}$ to
$0.003\gamma_{12}$, the resonance peaks will enhance. Obviously, a
small detuning is preferred in such a quantum dot system.

Figure 3 demonstrates the real and imaginary parts of the refractive
index $n$ and the ratio $-Re(n)/Im(n)$ as a function of the
tunneling with $\Delta/\gamma_{12}=0.001, 0.002, 0.003$. The other
parameters are the same as Fig.2. For the real applications of the
negative index materials, the ratio $-Re(n)/Im(n)$ has to be
seriously considered since the low-loss negative index materials are
desired. The curve of $Re(n)$ shows that the refractive index can be
adjusted and controlled
 from -5 to 3 continuously for $\Delta=0.001\gamma_{12}$. As $T_c$ is changed in $-0.22\gamma_{12}<T_c<-0.03\gamma_{12}$ and
$0.03\gamma_{12}<T_c<0.22\gamma_{12}$, the real part of refractive
index will be negative. The  ratio $-Re(n)/Im(n)$ curve shows that
the very low loss and the negative refractive index may be obtained
at some point near $T_c=0.18\gamma_{12}$ for
$\Delta=0.001\gamma_{12}$.

In conclusion, we have demonstrated the negative index of refraction
can be obtained by means of voltage-controlled the coupled quantum
dots system such as the vertically self-organized InAs quantum dots.
As the bias voltage is changed, the refractive index can be adjusted
continuously from negative to positive for some fixed detunings and
simultaneously the loss of light in the system will be small. The
SNIM and the DNIM can be achieved only by simply changing the
external bias voltage. We hope that our prediction will be observed
in the near future experiment.


This work has been supported in part by National Natural Science
Foundation of China (No.10774101) and the National Ministry of
Education Program for Training PhD.

\newpage
\section*{List of Figure Captions}

Fig. 1 (a) Schematic of the setup. A laser beam transmits from the
upper layer to the lower layer. The substrates and the top surface
are composed with transparent and conductive materials. (b) A
typical vertically coupled InAs/InGaAs asymmetrical quantum dots
system \cite{Qianghua}. (c) The energy levels of a quantum dot
system with the laser and the tunneling.

Fig. 2 The relative electric permittivity $\epsilon_r$ and the
relative magnetic permeability $\mu_r$ as a function of
$T_c/\gamma_{12}$ with three different detunings
($\Delta/\gamma_{12}=0.001, 0.002, 0.003$). The other parameters
used are $N=5\times10^{20}m^{-3}$, $\omega_{31}=0$, and
$\gamma_{32}=5\times10^{-4}\gamma_{12}$.

Fig. 3 $Re(n)$ and -$Re(n)/Im(n)$ of the system with three detunings
($\Delta/\gamma_{12}=0.001, 0.002, 0.003$). The other parameters are
the same as in Fig.2.


\clearpage
\begin{figure}[htbp]
\centering
\includegraphics[width=12cm]{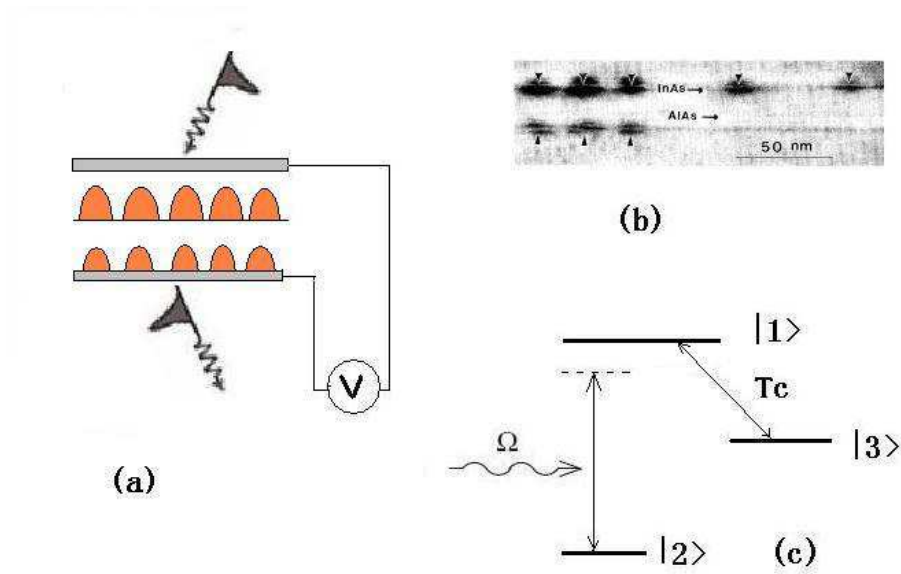}
\caption{(a) Schematic of the setup. A laser beam transmits from the
upper layer to the lower layer. The substrates and the top surface
are composed with transparent and conductive materials. (b) A
typical vertically coupled InAs/InGaAs asymmetrical quantum dots
system \cite{Qianghua}. (c) The energy levels of a quantum dot
system with the laser and the tunneling.}
\end{figure}
\clearpage
\begin{figure}[htbp]
\centering
\includegraphics[width=12cm]{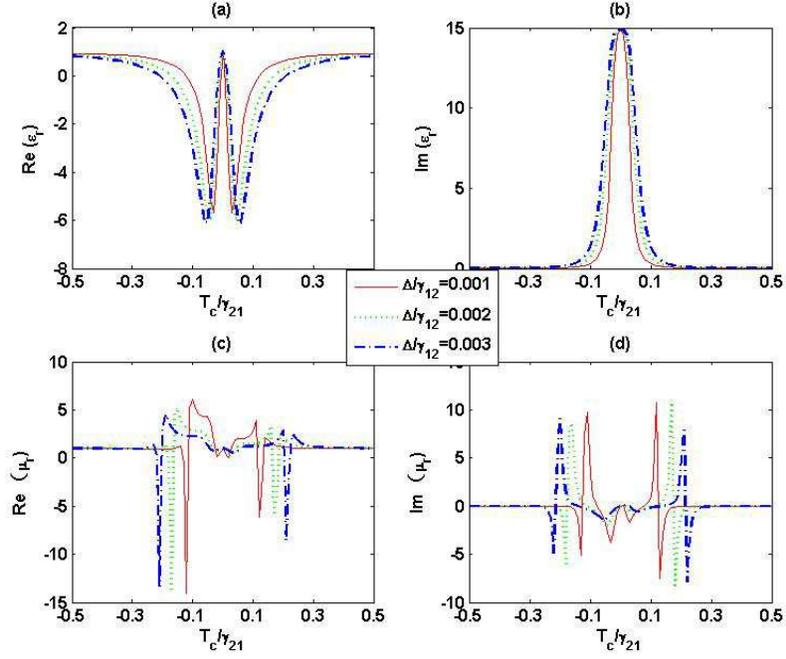}
\caption{The relative electric permittivity $\epsilon_r$ and the
relative magnetic permeability $\mu_r$ as a function of
$T_c/\gamma_{12}$ with three different detunings
($\Delta/\gamma_{12}=0.001,0.002,0.003$). The other parameters used
are $N=5\times10^{20}m^{-3}$, $\omega_{31}=0$ and
$\gamma_{32}=5\times10^{-4}\gamma_{12}$.}
\end{figure}
\clearpage
\begin{figure}[htbp]
\centering
\includegraphics[width=12cm]{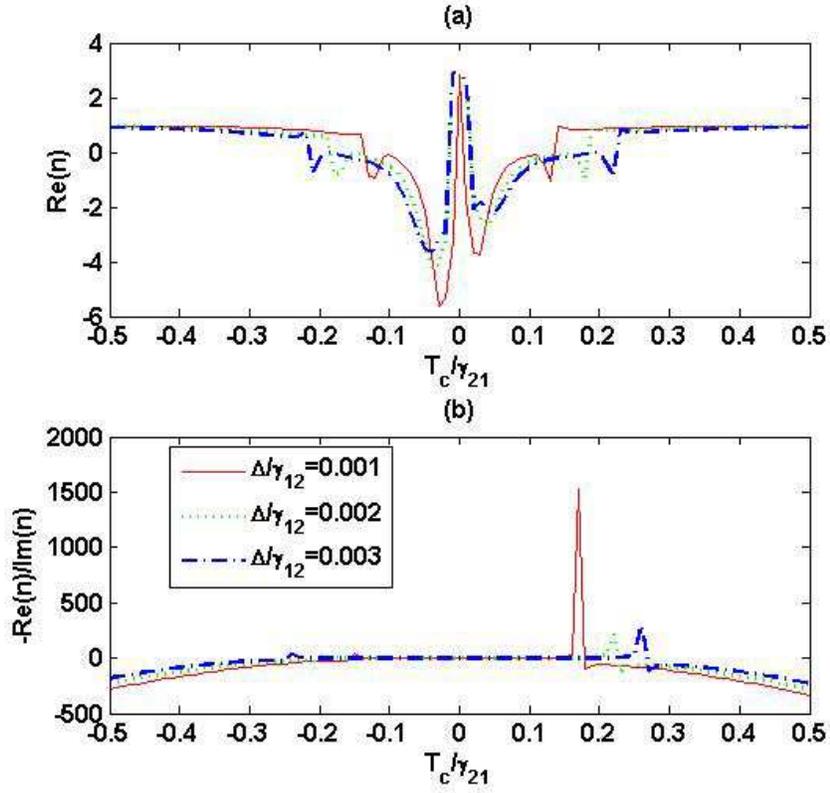}
\caption{$Re(n)$ and -$Re(n)/Im(n)$ of the system with three
detunings ($\Delta/\gamma_{12}=0.001,0.002,0.003$). The other
parameters are the same as in Fig.2.}
\end{figure}

\end{document}